# Grandma's local realistic theory for the Greenberger-Horne-Zeilinger experiment

BOON LEONG LAN

School of Engineering and Science, Monash University, 2 Jalan Kolej, 46150 Petaling Jaya, Selangor, Malaysia.

A realistic theory is constructed for the GHZ experiment. It is shown that the theory is local and it reproduces all the probabilistic predictions of quantum theory. This local realistic theory shows that GHZ had formulated Einstein's locality or no-action-at-a-distance principle incorrectly in their local realistic theory for the experiment.

In the GHZ experiment [1], the three photons (labeled 1, 2 and 3) produced are mathematically described by this state

$$|\psi\rangle = \frac{1}{\sqrt{2}}\left(|H\rangle_1|H\rangle_2|H\rangle_3 + |V\rangle_1|V\rangle_2|V\rangle_3\right) \quad (1)$$

where $|H\rangle_i$ is a horizontal linearly polarized state and $|V\rangle_i$ a vertical linearly polarized state. For each photon, either its linear polarization at 45º is measured (outcome is either *H'* or *V'* polarized) or its circular polarization is measured (outcome is either *L* or *R* polarized). The former measurement is called an *x* measurement and the latter is called a *y* measurement. In an experimental run, there are therefore eight possible system measurements: *yyx, yxy, xyy, yxx, xxy, xyx, xxx,* and *yyy*, where, for example, *yyx* means *y* measurement for photon 1, *y* measurement for photon 2 and *x* measurement for photon 3.

The three-photon GHZ state in Eq. (1) can be expressed in terms of the set of basis states for any one of the possible system measurements. For example, in terms of the basis states for the possible *yyx* system measurement,

$$|\psi\rangle = \frac{1}{2}\left(|R\rangle_1|L\rangle_2|H'\rangle_3 + |L\rangle_1|R\rangle_2|H'\rangle_3 + |R\rangle_1|R\rangle_2|V'\rangle_3 + |L\rangle_1|L\rangle_2|V'\rangle_3\right). \quad (2)$$

I will call each member of a set of basis states, which is of the form $|\ \rangle_1|\ \rangle_2|\ \rangle_3$, a system polarization state.

In Grandma's realistic theory for the GHZ experiment, in an experimental run, *prior to measurement*, the system of three photons is actually in one of the possible system polarization states for each of the eight *possible* system measurements. (In other words, in an experimental run, *prior to measurement*, the system of three photons is actually in eight system polarization states, one system polarization state for each *possible* system measurement.) For a possible system measurement, the probability that a particular system polarization state in the expansion for the GHZ state is the actual state of the system is given by the square of the amplitude of the expansion coefficient for that system polarization state. For example, for the possible *yyx* system measurement, the system of three photons is actually in one of the four possible system polarization states (see Eq. (2)): $|R\rangle_1|L\rangle_2|H'\rangle_3$ or $|L\rangle_1|R\rangle_2|H'\rangle_3$ or $|R\rangle_1|R\rangle_2|V'\rangle_3$ or $|L\rangle_1|L\rangle_2|V'\rangle_3$. The system is actually either in state $|R\rangle_1|L\rangle_2|H'\rangle_3$ with probability ¼, or in state $|L\rangle_1|R\rangle_2|H'\rangle_3$ with probability ¼, or in state $|R\rangle_1|R\rangle_2|V'\rangle_3$ with probability ¼, or in state $|L\rangle_1|L\rangle_2|V'\rangle_3$ probability ¼. Table 1 lists a possible set of eight *pre-existing* system polarization states, one state for each *possible* system measurement, that pre-existed before *any* system measurement is *chosen* in a hypothetical experimental run.

In an experimental run, the *chosen* system measurement reveals the corresponding *pre-existing* system polarization state. As an illustration, suppose, in an experimental run, the set of eight pre-existing system polarization states is given by Table 1. If the chosen system measurement is *yyx*, then the pre-existing system polarization state $|R\rangle_1|L\rangle_2|H'\rangle_3$ will be revealed by the measurement. However, if the chosen system measurement is *yyy*, then the pre-existing system polarization state $|L\rangle_1|L\rangle_2|R\rangle_3$ will be revealed by the measurement. Likewise, if the chosen system measurement is one of the other six possibilities, the corresponding pre-existing system polarization state listed in Table 1 will be revealed by the measurement.

Because the *chosen* system measurement merely reveals the corresponding *pre-existing* system polarization state, there is simply no action at a distance whatsoever. (In contrast, if each photon does not have a definite polarization before measurement, measurement on one photon compels that photon to acquire a definite polarization and instantaneously triggers the other two photons, spatially separated from the first, to also acquire a definite polarization each. This instantaneous 'triggering' is non-locality or action at a distance [2,3].) In other words, Grandma's realistic theory for the GHZ experiment is intrinsically local.

Moreover, for a chosen system measurement, since the probability that a particular system polarization state in the expansion for the GHZ state is the actual state of the system is given by the square of the amplitude of the expansion coefficient for that system polarization state, the probability of measuring the system in that state is given by the usual quantum mechanical probability. In other words, Grandma's local realistic theory reproduces all the probabilistic predictions of quantum theory for the GHZ experiment.

In Grandma's local realistic theory, each photon has eight *pre-existing* polarization states, one for each *possible* system measurement, which pre-existed before *any* system measurement is *chosen*. Measurement result (i.e., which of the eight pre-existing polarization states is revealed) for each photon therefore depends on which system measurement is *chosen*. However, this dependence of the measurement result for each photon on the chosen system measurement ***does not*** imply non-locality or action at a distance because the *chosen* system measurement merely reveals the corresponding *pre-existing* polarization state of each photon. This means that locality or no action at a distance ***does not*** require the measurement result for a photon not to depend on which measurements are chosen for the other two photons, contrary to what GHZ [1,4,5] had assumed in their local realistic theory. In other words, GHZ [1,4,5] had made a mistake in formulating Einstein's locality or no-action-at-a-distance principle in their local realistic theory.

A local realistic theory has also been constructed [6,7] for the Einstein-Podolsky-Rosen-Bohm (EPRB) experiment that reproduces all the predictions of quantum theory. The theory shows that Bell [8], Mermin [2,9], and Hess and Philipp [10-12] had also formulated Einstein's locality or no-action-at-a-distance principle incorrectly in their local realistic theories for the EPRB experiment. The mis-formulation by GHZ is the same as the mis-formulation by Bell, Mermin, and Hess and Philipp.

| Possible system measurement | Pre-existing system polarization state |
|---|---|
| yyx | $\|R\rangle_1\|L\rangle_2\|H'\rangle_3$ |
| yyy | $\|L\rangle_1\|L\rangle_2\|R\rangle_3$ |
| yxx | $\|L\rangle_1\|V'\rangle_2\|H'\rangle_3$ |
| yxy | $\|L\rangle_1\|H'\rangle_2\|R\rangle_3$ |
| xyy | $\|H'\rangle_1\|R\rangle_2\|L\rangle_3$ |
| xxx | $\|V'\rangle_1\|V'\rangle_2\|H'\rangle_3$ |
| xxy | $\|V'\rangle_1\|H'\rangle_2\|L\rangle_3$ |
| xyx | $\|H'\rangle_1\|L\rangle_2\|V'\rangle_3$ |

**Table 1**   Pre-existing system polarization states, one state for each possible system measurement, in a hypothetical experimental run.